\documentstyle[12pt]{article} \voffset=0mm \headheight=0mm
\date{}
\author{Valerii Dryuma\thanks{Work supported in part by MURST, Italy}\\[5mm]
{\it Institute of Mathematics and Informatics, AS RM,}\\[3mm]
{\it 5 Academiei Street, 2028 Chisinau, Moldova},\\[3mm]{\it e-mail:
valery@dryuma.com; cainar@mail.md; dryuma@math.md} }
\title{ON THE RIEMANN EXTENSION OF THE SCHWARZSCHILD METRICS}

\begin{document}
\maketitle
\date{}
\maketitle
\begin{abstract}
   Some solutions of the Einstein equations for the 8-dimensional Riemann extension of the classical
   Schwarzschild 4-dimensional metrics are considered.
\end{abstract}
\maketitle

\section{Introduction}

   The notion of the Riemann extension of nonriemannian spaces were  first introduced in ~(\cite{dryuma:paterson&walker}).
Main idea of this theory is  application of the methods of Riemann geometry for studying of the properties
of nonriemannian spaces.

For example the system differential equations in form
\begin{equation} \label{dryuma:eq1}
\frac{d^2 x^{k}}{ds^2}+\Pi^k_{ij}\frac{dx^{i}}{ds}\frac{dx^{j}}{ds}=0
\end{equation}
with arbitrary coefficients $\Pi^k_{ij}(x^l)$ can be considered as the system of geodesic equations of
affinely connected space with local coordinates $x^k$.

 For the n-dimensional Riemannian spaces with the metrics
$$
{^n}ds^2=g_{ij}dx^i dx^j
$$
the system of geodesic equations looks same but the coefficients $\Pi^k_{ij}(x^l)$ now have very special form and
depends from the choice of the metric $g_{ij}$.
\[
\Pi^i_{kl}=\Gamma^i_{kl}=\frac{1}{2}g^{im}(g_{mk,l}+g_{ml,k}-g_{kl,m})
\]

In order that the methods of Riemann geometry can be applied for studying of the properties of the spaces with
equations (\ref{dryuma:eq1}) the construction of 2n-dimensional extension of the space with local coordinates $x^i$  was introduced .

The metric of extended space constructs with help of coefficients of equation (\ref{dryuma:eq1}) and looks as follows
\begin{equation} \label{dryuma:eq2}
{^{2n}}ds^2=-2\Pi^k_{ij}(x^l)\Psi_k dx^i dx^j+2d \Psi_k dx^k
\end{equation}
where $\Psi_{k}$ are the coordinates of additional space.

The important property of such type metric is that the geodesic
 equations of metric (\ref{dryuma:eq2})  decomposes into two parts
\begin{equation} \label{dryuma:eq3}
\ddot x^k +\Gamma^k_{ij}\dot x^i \dot x^j=0,
\end{equation}
and
\begin{equation} \label{dryuma:eq4}
\frac{\delta^2 \Psi_k}{ds^2}+R^l_{kji}\dot x^j \dot x^i \Psi_l=0,
\end{equation}
where
\[
\frac{\delta \Psi_k}{ds}=\frac{d \Psi_k}{ds}-\Pi^l_{jk}\Psi_l\frac{d x^j}{ds}.
\]

The first part (\ref{dryuma:eq3}) of complete system is the system of equations for geodesics of basic space with local
coordinates $x^i$ and they does not contains the coordinates $\Psi_k$.

 The second part (\ref{dryuma:eq4}) of system of geodesic equations  has the form
of linear $4\times4$ matrix system of second order ODE's for coordinates $\Psi_k$
\begin{equation} \label{dryuma:eq5}
\frac{d^2 \vec \Psi}{ds^2}+A(s)\frac{d \vec \Psi}{ds}+B(s)\vec \Psi=0.
\end{equation}

From this point of view we have the case of geodesic extension of basic space in local coordinates $(x^i)$.

It is important to note that the geometry of extended space is connected with geometry of basic space.
For example the property of this space to be Ricci-flat keeps also for the extended space.

This fact give us the possibility to use the linear system of equation (\ref{dryuma:eq5}) for studying
of the properties of basic space.

In particular the invariants of the $4\times4$ matrix-function
\[
E=B-\frac{1}{2}\frac{d A}{ds}-\frac{1}{4}A^2
\]
under change of the coordinates $\Psi_k$ can be used for that.

 The first applications of the notion of extended spaces the studying of nonlinear second order differential
 equations connected with nonlinear dynamical systems was done in works of author
 ~(\cite{dryuma3:dryuma,dryuma4:dryuma,dryuma5:dryuma}).

 Here we consider the properties of extended spaces for the Schwarzschild metrics in General Relativity
  ~(\cite{dryuma1:dryuma,dryuma2:dryuma}).

\section{The Schwarzschild space-time and geodesic equation}.

The line element of standard metric of the Schwarzschild space-time
in coordinate  system $x ,\theta, \phi,t$ has the form

\begin{equation} \label{dryuma:eq6}
ds^2=-\frac{1}{(1-2M/x)}dx^2-x^2(d\theta^2+\sin^2 \theta d\phi^2)+
(1-2M/x)dt^2.
\end{equation}

The geodesic equations of this type of the metric are
\begin{equation} \label{dryuma:eq7}
{\frac {d^{2}}{d{s}^{2}}}x(s)-{\frac {M\left ({\frac {d}{ds}}x(s)
\right )^{2}}{\left (x-2\,M\right )x}}+\left (-x+2\,M\right )\left ({
\frac {d}{ds}}\theta(s)\right )^{2}-
\end{equation}
\[-\left (x-2\,M\right )\left (\sin(
\theta)\right )^{2}\left ({\frac {d}{ds}}\phi(s)\right )^{2}+\\
{\frac {
\left (x-2\,M\right )M\left ({\frac {d}{ds}}t(s)\right )^{2}}{{x}^{3}}
}=0,
\]
\begin{equation} \label{dryuma:eq8}
{\frac {d^{2}}{d{s}^{2}}}\theta(s)+2\,{\frac {\left ({\frac {d}{ds}}x(
s)\right ){\frac {d}{ds}}\theta(s)}{x}}-\sin(\theta)\cos(\theta)\left
({\frac {d}{ds}}\phi(s)\right )^{2}=0,
\end{equation}
\begin{equation}\label{dryuma:eq9}
 {\frac {d^{2}}{d{s}^{2}}}\phi(s)+2\,{\frac {\left ({\frac {d}{ds}}x(s)
\right ){\frac {d}{ds}}\phi(s)}{x}}+2\,{\frac {\cos(\theta)\left ({
\frac {d}{ds}}\theta(s)\right ){\frac {d}{ds}}\phi(s)}{\sin(\theta)}}=0,
\end{equation}
\begin{equation} \label{dryuma:eq10}
{\frac {d^{2}}{d{s}^{2}}}t(s)-2\,{\frac {M\left ({\frac {d}{ds}}x(s)
\right ){\frac {d}{ds}}t(s)}{x\left (2\,M-x\right )}}=0.
\end{equation}

The symbols of Christoffel of the metric (\ref{dryuma:eq6}) looks as
\[
\Gamma^1_{11}=\frac{M}{x(2M-x)},\quad
\Gamma^1_{22}=(2M-x),\quad \Gamma^1_{33}=(2M-x)\sin^2 \theta,\]\[
\quad \Gamma^1_{44}=-\frac{M(2M-x)}{x^3},
 \Gamma^2_{12}=\frac{1}{x},\quad \Gamma^2_{33}=-\sin \theta \cos \theta,\]\[
\quad \Gamma^3_{23}=\frac{\cos \theta}{\sin \theta},\quad
\Gamma^4_{14}=-\frac{M}{x(2M-x)},\quad \Gamma^3_{13}=\frac{1}{x}.
\]

The equations of geodesic  (\ref{dryuma:eq7}--\ref{dryuma:eq10}) have the
first integrals
\begin{equation}\label{dryuma:eq11}
{\frac {d}{ds}}x(s)=h\sqrt {1-\left (1-2\,{\frac {M}{x(s)}}\right )
\left ({h}^{-2}+{\frac {{C}^{2}}{\left (x(s)\right )^{2}}}\right )},
\end{equation}
\[\left ({\frac {d}{ds}}\theta(s)\right )^{2}={h}^{2}\left ({B}^{2}-{
\frac {{C}^{2}}{\left (\sin(\theta)\right )^{2}}}\right ){x}^{-4},
\]
\begin{equation}\label{dryuma:eq12}
{\frac {d}{ds}}\phi(s)={\frac {hC}{{x}^{2}\left (\sin(\theta)\right )^
{2}}}
\end{equation}
\[{\frac {d}{ds}}t(s)=h\left (1-2\,{\frac {M}{x(s)}}\right )^{-1},
\]
where a dot denotes differentiation with respect to parameter $s$ and
 $(C, B, h)$ are the constants of motion.

\section{ The Riemann extension of the Schwarzschild metric}

Now with help of the formulae (\ref{dryuma:eq2}) we construct the eight-dimensional
extension of basic metric (\ref{dryuma:eq6})
\begin{equation}\label{dryuma:eq13}
ds^2=-\frac{2M}{x(2M-x)}Pdx^2-\frac{2}{x}Qdxd\theta-2(2M-x)Pd\theta^2-
\frac{2}{x}Udxd\phi+
\end{equation}
+\[2\frac{M}{x(2M-x)}Vdxdt-
2\frac{\cos\theta}{\sin\theta}Ud\phi d\theta-
2((2M-x)\sin^2\theta P-\sin\theta \cos\theta Q)d\phi^2+\]\[
+2\frac{M(2M-x)}{x^3}Pdt^2+2dxdP+2d\theta dQ+2d\phi dU+2dtdV,
\]
where $(P,Q,U,V)$ are the additional coordinates of extension.

   The metrics of a given type are the metrics with vanishing curvature invariants. They play an
important role in general theory of Riemannian spaces. In particular the metrics for pp-waves
in General Relativity belong to this class.

    The eight-dimensional space in local coordinates $(x,\theta,\phi,t,P,Q,U,V)$
 with this type of metric is also the Einstein space with condition on the Ricci tensor
\[
{^8}R_{ik}=0.
\]

  The complete system of  geodesic equations for the metric (\ref{dryuma:eq13})  decomposes into two groups of
equations.

The first group coincides with the equations (\ref{dryuma:eq7}-\ref{dryuma:eq10})
on the coordinates $(x,\theta,\phi,t)$ and second  part forms the linear system of equations for coordinates
$P,Q,U,V$.

 They are defined as
\[
\ddot P+\frac{2M}{x(x-2M)}\dot x \dot P-\frac{2}{x}\dot \theta \dot Q -
\frac{2}{x}\dot \phi \dot U-\frac{2M}{x(x-2M)} \dot t \dot V -
\]
\[
-\left(\frac{2M}{x^2(x-2M)}\dot x^2+
\frac{(x-2M)}{x}\dot \theta^2+\frac{\sin^2 \theta(x-2M)}{x}\dot \phi^2+
\frac{2M(x-2M)}{x^4}\dot t^2\right)P+
\]
\[
+\left(\frac{4}{x^2}\dot x \dot \theta-\frac{2\cos \theta}{x}\dot \phi^2\right)Q+
\left(\frac{4}{x^2}\dot x \dot \phi+\frac{4\cos \theta}{x\sin \theta}\dot \theta \dot \phi\right)U+
\left(\frac{4M^2}{x^2(x-2m)^2}\dot x \dot t\right)V=0,
\]

\[
\ddot Q+2(x-2m)\dot \theta \dot P-\frac{2}{x}\dot x \dot Q -
\frac{2\cos \theta}{\sin \theta}\dot \phi \dot U-
\frac{2(x-4M)}{x}\dot x \dot \theta P+
\]
\[
+\left(\frac{2(x-3M)}{x^2(x-2M)}\dot x^2-\frac{2(x-2M)}{x}\dot \theta^2-\frac{(x-4M \sin^2 \theta)}{x} \dot \phi^2+
\frac{2M(x-2M)}{x^4}\dot t^2 \right)Q+\]\[+\left(\frac{4\cos \theta}{x\sin \theta}\dot x \dot \phi+
\frac{4\cos^2 \theta}{\sin^2 \theta}\dot \theta \dot \phi\right)U=0,
\]

\[
\ddot U+2\sin^2 \theta(x-2M)\dot \phi \dot P+2\sin \theta \cos \theta\dot \phi \dot Q -
\left(\frac{2\cos \theta}{\sin \theta}\dot \theta +\frac{2}{x}\dot x \right)\dot U-
\frac{2\sin^2 \theta(x-4M)}{x}\dot x \dot \phi P-
\]
\[
-\left(\frac{4\sin \theta \cos \theta}{x}\dot x \dot \phi+2\dot \theta \dot \phi \right)Q+\]
\[
+\left(\frac{2(x-3M)}{x^2(x-2m)}\dot x^2+\frac{4\cos \theta}{x\sin \theta}\dot x \dot \theta +
\frac{2(x\cos^2 \theta+2M\sin^2 \theta)}{x\sin^2 \theta}\dot \theta^2 \right)-
\]
\[
-\left(\frac{2(x-2M\sin^2\theta)}{x}\dot \phi^2\right)U=0,
\]

\[
\ddot V-\frac{2M(x-2m)}{x^3}\dot t \dot P-\frac{2M}{x(x-2M)}\dot x \dot V+
\frac{4M(x-2M)}{x^4}\dot x \dot t P+\]\[
+\left(\frac{2M(2x-3M)}{x^2(x-2M)^2}\dot x^2-\frac{2M}{x}\dot \theta^2-\frac{2M\sin^2 \theta}{x}\dot \phi^2+\frac{2M^2}{x^4}\dot t^2\right)V=0.
\]

So we get the linear  matrix-second order ODE for the coordinates $U,V,P,Q$
\begin{equation}\label{dryuma:eq14}
\frac{d^2\Psi}{ds^2}+A(x,\theta,\phi,t)\frac{d\Psi}{ds}+B(x,\theta,\phi,t)\Psi=0,
\end{equation}
where
\[
\Psi(s)=\left(\begin{array}{c}
P(s)\\
Q(s)\\
U(s)\\
V(s)
\end{array}\right)
\]
and $A,B$ are some $4 \times 4$ matrix-functions depending from the coordinates
$x^i(s)$ and their derivatives.

We shall study this system of equations at the condition $\theta=\pi/2$.

In this case we get the  system for the coordinates of basic space
\[
{\frac {d^{2}}{d{s}^{2}}}x(s)+{\frac {M\left ({\frac {d}{ds}}x(s)
\right )^{2}}{x(s)\left (2\,M-x(s)\right )}}+\left (2\,M-x(s)\right )
\left ({\frac {d}{ds}}\phi(s)\right )^{2}-\]\[-{\frac {M\left (2\,M-x(s)
\right )\left ({\frac {d}{ds}}t(s)\right )^{2}}{\left (x(s)\right )^{3
}}}=0,
\]
\[
{\frac {d^{2}}{d{s}^{2}}}\phi(s)+2\,{\frac {\left ({\frac {d}{ds}}x(s)
\right ){\frac {d}{ds}}\phi(s)}{x(s)}}=0
\]
\[
{\frac {d^{2}}{d{s}^{2}}}t(s)-\,{\frac {M\left ({\frac {d}{ds}}x(s)
\right ){\frac {d}{ds}}t(s)}{x(s)\left (2\,M-x(s)\right )}}=0
\]

and the system of equations for the supplementary coordinates
\[
\left (2\,{\frac {M\left ({\frac {d}{ds}}x(s)\right )^{2}}{\left (2\,M
-x(s)\right )\left (x(s)\right )^{2}}}-{\frac {\left (-2\,M+x(s)
\right )\left ({\frac {d}{ds}}\phi(s)\right )^{2}}{x(s)}}\right)P(s)+\]\[+\left(2\,{\frac {M
\left (2\,M-x(s)\right )\left ({\frac {d}{ds}}t(s)\right )^{2}}{\left
(x(s)\right )^{4}}}\right )P(s)+\]\[+{\frac {d^{2}}{d{s}^{2}}}P(s)+
4\,{
\frac {U(s)\left ({\frac {d}{ds}}x(s)\right ){\frac {d}{ds}}\phi(s)}{
\left (x(s)\right )^{2}}}+4\,{\frac {{M}^{2}V(s)\left ({\frac {d}{ds}}
x(s)\right ){\frac {d}{ds}}t(s)}{\left (x(s)\right )^{2}\left (2\,M-x(
s)\right )^{2}}}-2\,{\frac {M\left ({\frac {d}{ds}}x(s)\right ){\frac
{d}{ds}}P(s)}{x(s)\left (2\,M-x(s)\right )}}-\]\[-2\,{\frac {\left ({\frac
{d}{ds}}\phi(s)\right ){\frac {d}{ds}}U(s)}{x(s)}}+2\,{\frac {M\left (
{\frac {d}{ds}}t(s)\right ){\frac {d}{ds}}V(s)}{x(s)\left (2\,M-x(s)
\right )}}=0,
\]

\[
\left (2\,{\frac {\left (-x(s)+3\,M\right )\left ({\frac {d}{ds}}x(s)
\right )^{2}}{\left (2\,M-x(s)\right )\left (x(s)\right )^{2}}}-{
\frac {\left (x(s)-4\,M\right )\left ({\frac {d}{ds}}\phi(s)\right )^{
2}}{x(s)}}\right)Q(s)-\]\[-\left(2\,{\frac {M\left (2\,M-x(s)\right )\left ({\frac {d}{ds}}t
(s)\right )^{2}}{\left (x(s)\right )^{4}}}\right )Q(s)+{\frac {d^{2}}{
d{s}^{2}}}Q(s)-
2\,{\frac {\left ({\frac {d}{ds}}x(s)\right ){\frac {d}
{ds}}Q(s)}{x(s)}}=0,
\]
\[
-2\,{\frac {\left (x(s)-4\,M\right )\left ({\frac {d}{ds}}x(s)\right )
\left ({\frac {d}{ds}}\phi(s)\right )P(s)}{x(s)}}+
\]
\[
+2\,\left (-{\frac {
\left (-2\,M+x(s)\right )\left ({\frac {d}{ds}}\phi(s)\right )^{2}}{x(
s)}}+{\frac {\left (-x(s)+3\,M\right )\left ({\frac {d}{ds}}x(s)
\right )^{2}}{\left (2\,M-x(s)\right )\left (x(s)\right )^{2}}}\right)U(s)-\]\[\left({
\frac {M\left (2\,M-x(s)\right )\left ({\frac {d}{ds}}t(s)\right )^{2}
}{\left (x(s)\right )^{4}}}\right )U(s)+
{\frac {d^{2}}{d{s}^{2}}}U(s)-
2\,\left (2\,M-x(s)\right )\left ({\frac {d}{ds}}\phi(s)\right ){
\frac {d}{ds}}P(s)-\]\[-2\,{\frac {\left ({\frac {d}{ds}}x(s)\right ){
\frac {d}{ds}}U(s)}{x(s)}}=0,
\]

\[
\left (-2\,{\frac {M\left (3\,M-2\,x(s)\right )\left ({\frac {d}{ds}}x
(s)\right )^{2}}{\left (x(s)\right )^{2}\left (2\,M-x(s)\right )^{2}}}
-2\,{\frac {M\left ({\frac {d}{ds}}\phi(s)\right )^{2}}{x(s)}}+2\,{
\frac {{M}^{2}\left ({\frac {d}{ds}}t(s)\right )^{2}}{\left (x(s)
\right )^{4}}}\right )V(s)+
\]
\[
+{\frac {d^{2}}{d{s}^{2}}}V(s)-4\,{\frac {MP
(s)\left (2\,M-x(s)\right )\left ({\frac {d}{ds}}x(s)\right ){\frac {d
}{ds}}t(s)}{\left (x(s)\right )^{4}}}+2\,{\frac {M\left ({\frac {d}{ds
}}x(s)\right ){\frac {d}{ds}}V(s)}{x(s)\left (2\,M-x(s)\right )}}+\]\[+2\,{
\frac {M\left (2\,M-x(s)\right )\left ({\frac {d}{ds}}t(s)\right ){
\frac {d}{ds}}P(s)}{\left (x(s)\right )^{3}}}=0.
\]

In this case the matrix $A$ takes the form
\[
A=- \left [\begin {array}{cccc} -2\,{\frac {\left ({\frac {d}{ds}}x(s)
\right )M}{x(s)\left (-2\,M+x(s)\right )}}&0&2\,{\frac {{\frac {d}{ds}
}\phi(s)}{x(s)}}&2\,{\frac {M{\frac {d}{ds}}t(s)}{x(s)\left (-2\,M+x(s
)\right )}}\\\noalign{\medskip}0&2\,{\frac {{\frac {d}{ds}}x(s)}{x(s)}
}&0&0\\\noalign{\medskip}-2\,\left (-2\,M+x(s)\right ){\frac {d}{ds}}
\phi(s)&0&2\,{\frac {{\frac {d}{ds}}x(s)}{x(s)}}&0\\\noalign{\medskip}
2\,{\frac {\left (-2\,M+x(s)\right )\left ({\frac {d}{ds}}t(s)\right )
M}{\left (x(s)\right )^{3}}}&0&0&2\,{\frac {\left ({\frac {d}{ds}}x(s)
\right )M}{x(s)\left (-2\,M+x(s)\right )}}\end {array}\right ],
\]

and matrix $B$ has an elements
\[
B_{11}=\]\[{\frac {\left ({\frac {d}{ds}}\phi(s)
\right )^{2}\left (x(s)\right )^{4}(x(s)-4\,M)+4\,{M}^{2}\left ({\frac {d}{ds}}
\phi(s)\right )^{2}\left (x(s)\right )^{3}+2\,M\left ({\frac {d}{ds}}t
(s)\right )^{2}\left (x(s)\right )^{2}}{
\left (-2\,M+x(s)\right )\left (x(s)\right )^{4}}}+
\]
\[
+{\frac{2\,M\left ({\frac {d}{ds}}x(s)
\right )^{2}\left (x(s)\right )^{2}\left (-2\,M+x(s)\right )\left (x(s)\right )^{4}-
8\,{M}^{2}\left ({\frac {d}{ds}}t(
s)\right )^{2}x(s)+8\,{M}^{3}\left ({\frac {d}{ds}}t(s)\right )^{2}}
{\left (-2\,M+x(s)\right )\left (x(s)\right )^{4}}},
\]
\[
B_{12}=0, B_{13}=-4\,{\frac {
\left ({\frac {d}{ds}}x(s)\right ){\frac {d}{ds}}\phi(s)}{\left (x(s)
\right )^{2}}}, B_{14}=-4\,{\frac {{M}^{2}\left ({\frac {d}{ds}}x(s)\right ){
\frac {d}{ds}}t(s)}{\left (x(s)\right )^{2}\left (4\,{M}^{2}-4\,x(s)M+
\left (x(s)\right )^{2}\right )}},
\]
\[
B_{21}=0,
\quad B_{22}=\]\[{\frac {-8\,{M}
^{3}\left ({\frac {d}{ds}}t(s)\right )^{2}+\left ({\frac {d}{ds}}\phi(
s)\right )^{2}\left (x(s)\right )^{5}+8\,{M}^{2}\left ({\frac {d}{ds}}
\phi(s)\right )^{2}\left (x(s)\right )^{3}-2\,\left ({\frac {d}{ds}}x(
s)\right )^{2}\left (x(s)\right )^{3}}{\left (-2\,M+x(s)\right )\left (x(s)\right )^{4}}}-
\]
\[
-{\frac{2\,M\left ({\frac {d}{ds}}t(s)
\right )^{2}\left (x(s)\right )(x(s)+4\,M)+6\,M\left ({\frac {d}{ds}}x(s)\right )^{2}\left (x(
s)\right )^{2}-6\,M\left ({\frac {d}{ds}}\phi(s)\right )^{2}\left (x(s
)\right )^{4}}{\left (-2\,M+x(s)\right )\left (x(s)\right )^{4}}},\]
\[
\quad B_{23}=0,\quad B_{24}=0,
\]
\[
B_{31}=2\,{\frac {\left (x(s)-4\,M\right )\left ({\frac {d}{ds}}x(s)\right ){\frac {d}{ds}}\phi(s)}{x(s)}},\quad B_{32}=0,\]
\[
 B_{33}=\]\[-2\,{\frac {-
\left ({\frac {d}{ds}}\phi(s)\right )^{2}\left (x(s)\right )^{3}(x(s)^2+4M^2)+4\,{M
}^{3}\left ({\frac {d}{ds}}t(s)\right )^{2}-3\,M\left ({\frac {d}{ds}}
x(s)\right )^{2}\left (x(s)\right )^{2}}{\left (-2\,M+x(s)\right )\left (x(s)\right )^{4}}}+
\]
\[
+{\frac{\left ({\frac {d}{ds}}x(s
)\right )^{2}\left (x(s)\right )^{3}+4\,M\left ({\frac {d}{ds}}\phi(s)
\right )^{2}\left (x(s)\right )^{4}-4\,{M}^{2}\left ({\frac {d}{ds}}t(
s)\right )^{2}x(s)+M\left ({\frac {d}{ds}}t(s)\right )^{2}\left (x(s)
\right )^{2}}{\left (-2\,M+x(s)\right )\left (x(s)\right )^{4}}},\]\[ B_{34}=0,\quad
B_{41}=-4\,{\frac {\left (-2\,M+x(s)\right )\left ({
\frac {d}{ds}}t(s)\right )\left ({\frac {d}{ds}}x(s)\right )M}{\left (
x(s)\right )^{4}}},\quad
B_{42}=0,\quad B_{43}=0,\quad
\]
\[
B_{44}=-2\,{\frac {M\left (-3\,M\left ({\frac {d}{ds}}
x(s)\right )^{2}\left (x(s)\right )^{2}-\left ({\frac {d}{ds}}\phi(s)
\right )^{2}\left (x(s)\right )^{5}-4\,{M}^{2}\left ({\frac {d}{ds}}
\phi(s)\right )^{2}\left (x(s)\right )^{3}\right)}{\left (x(s)\right )^{4}\left (4\,{M}^{2}-4\,x(s)M+\left (x(s)\right
)^{2}\right )}}-
\]
\[
-2\,{\frac{M\left(4\,M\left ({\frac {d}{ds}}
\phi(s)\right )^{2}\left (x(s)\right )^{4}+2\,\left ({\frac {d}{ds}}x(
s)\right )^{2}\left (x(s)\right )^{3}+M\left ({\frac {d}{ds}}
t(s)\right )^{2}(2M- x(s))^2\right)}
{\left (x(s)\right )^{4}\left (4\,{M}^{2}-4\,x(s)M+\left (x(s)\right
)^{2}\right )}}.
\]

 Now we will integrate our system.

 Remark that the equation for the coordinate $Q(s)$ is independent from others equations and can be reduced after
 the substitution
\[
Q(s)=x(s)F(s)
\]
to the  equation for the function $F(s)$

\begin{equation}\label{dryuma:eq15}
{\frac {d^{2}}{d{s}^{2}}}F(s)+\left({\frac {M}{\left (x(s)\right )^{3}}}
+3\,{\frac {{C}^{2}{h}^{2}M}{\left (x(s)\right )^{5}}}\right)F(s)=0.
\end{equation}

    To integrate the equations for the coordinates $P(s)$, $U(s)$, $V(s)$ we use the relation
\begin{equation}\label{dryuma:eq16}
\left ({\frac {d}{ds}}x(s)\right )P(s)+\left ({\frac {d}{ds}}\theta(s)
\right )Q(s)+\left ({\frac {d}{ds}}\phi(s)\right )U(s)+\left ({\frac {
d}{ds}}t(s)\right )V(s)-
\end{equation}
\[
-1/2\,s-\mu=0
\]

which is consequence of the well known first integral of geodesic equations of arbitrary Riemann space
\[
g_{ik}\frac {dx^i(s)}{ds} \frac {dx^k(s)}{ds}=const.
\]

     In our case it takes the form

\[
\left ({\frac {d}{ds}}x(s)\right )P(s)+\left ({\frac {d}{ds}}\phi(s)\right )U(s)+\left ({\frac {
d}{ds}}t(s)\right )V(s)-1/2\,s-\mu=0.
\]

Solving this equation with respect the function $V(s)$
\begin{equation}\label{dryuma:eq17}
V(s)=-1/2\,{\frac {2\,\left ({\frac {d}{ds}}x(s)\right
)P(s)+2\,\left ({ \frac {d}{ds}}\phi(s)\right
)U(s)-s-2\,\mu}{{\frac {d}{ds}}t(s)}}
\end{equation}
and substituting  this expression into the last two equations of
the system we get the following two equations for coordinates
$U(s)$
\[
\frac{d^2}{ds^2}U(s)=
 -2\,{\frac {\left (-2\,Ch\left (x(s)\right )^{3}M+Ch\left
(x(s)\right )^{4}\right ){\frac {d}{ds}}P(s)}{\left (x(s)\right
)^{5}}}+
\]
\[
+2\sqrt {{\frac {\left (x(s)\right )^{3}{h}^{2}-\left
(x(s)\right )^{3}-{C}^{2 }{h}^{2}x(s)+2\,M\left (x(s)\right)^{2}+
2\,{C}^{2}{h}^{2}M}{\left (x(s)\right )^{3}{h}^{2}}}}\]\[ \left( h {\frac
{d}{ds}}U(s)\left (x(s)\right )^{-1}-
{h}^{2} C \left (x(s) \right )^{2}\left(4\,M-x(s)\right)P(s)\left (x(s)\right )^{-5}\right)-
\]
\[
-2\,{\frac {\left
(3\, M\left (x(s)\right )^{2}-\left (x(s)\right )^{3}+\left
(x(s)\right )^{
3}{h}^{2}-2\,{C}^{2}{h}^{2}x(s)+5\,{C}^{2}{h}^{2}M\right
)U(s)}{\left (x(s)\right )^{5}}}
\]
and $P(s)$
\[
\frac{d^2}{ds^2}P(s)+\frac{4\,Mh}{x(s)(x(s)-2M)}\sqrt {{\frac {\left (x(s)\right
)^{3}\left({h}^{2}-1\right)-{C}^{2}{h}^{2}x(s)+2\,M\left
(x(s)\right )^{2}+2\,{C}^{2}{h}^{2} M}{\left (x(s)\right
)^{3}{h}^{2}}}}\times \]\[\times {\frac {d}{ds}}P(s)=\]\[
-{\frac {\left (-2\,\left ( x(s)\right )^{3}Ch+6\,\left (x(s)\right
)^{2}ChM\right ){\frac {d}{ds} }U(s)}{\left (x(s)\right )^{5}\left
(-2\,M+x(s)\right )}}-
\]
\[
-{\frac { \left (2\,\left (x(s)\right )^{3}M(1-2\,{h}^{2})- \left
(x(s)\right ){h}^{2}{C}^{2}(x(s)-8M)-6\,\left (x(s)\right
)^{2}{M}^{2 }\right)P(s)}{\left (x(s)\right )^{5}\left
(-2\,M+x(s)\right
)}}-\]\[
-{\frac{14\,{h}^{2}{C}^{2}{M}^{2}P(s)}{\left
(x(s)\right )^{5}\left (-2\,M+x(s)\right )}}-\]
\[-{\sqrt {\frac
{\left (x(s)\right )^{3}{h}^{2}- \left (x(s)\right
)^{3}-{C}^{2}{h}^{2}x(s)+2\,M\left (x(s)\right )^{2}
+2\,{C}^{2}{h}^{2}M}{\left (x(s)\right
)^{3}{h}^{2}}}}\times\]\[\times \frac{\left (4\,\left (x(s) \right
)^{2}{h}^{2}C-12\,x(s){h}^{2 }CM\right )U(s)}{\left (x(s)\right
)^{5 }\left (-2\,M+x(s)\right)}+{\frac {M}{x(s)\left (-2\,M+x(s)
\right )}}
\]

    So we have showed that  every motion on orbit in usual space corresponds the motion in  additional space.

Let us consider some examples.

According to  ~(\cite{dryuma4:bogorod}) in the Schwarzshild space-time exists the cyclic orbit

\[
x(s)=6M
\]
which is the solution of the geodesic equations at the condition
\[
h=1,\quad C=3\sqrt(2)M.
\]

 In this case  our system takes the form
\begin{equation}\label{dryuma:eq18}
{\frac {d^{2}}{d{s}^{2}}}P(s)-1/48\,{\frac {\sqrt {2}{\frac {d}{ds}}U(
s)}{{M}^{2}}}-{\frac {19}{864}}\,{\frac {P(s)}{{M}^{2}}}-1/24\,{M}^{-1}=0,
\end{equation}
and
\begin{equation}\label{dryuma:eq19}
{\frac {d^{2}}{d{s}^{2}}}U(s)+2/3\,\sqrt {2}{\frac {d}{ds}}P(s)-{
\frac {1}{216}}\,{\frac {U(s)}{{M}^{2}}}=0.
\end{equation}

   The simplest solution of this system looks as
\begin{equation}\label{dryuma:eq20}
P(s)=-{\frac {36}{19}}\,M+A\sin\left({\frac {1}{72}}\,{\frac {\sqrt {3+3\,i
\sqrt {303}}s}{M}}\right)
\end{equation}
and
\begin{equation}\label{dryuma:eq21}
U(s)=A\left (39+i\sqrt {303}\right )\sqrt {2}M\cos\left({\frac {1}{72}}\,{
\frac {\sqrt {3+3\,i\sqrt {303}}s}{M}}\right){\frac {1}{\sqrt {3+3\,i\sqrt {
303}}}}
\end{equation}
where $A$ is arbitrary parameter.

The equation for coordinate $Q(s)$ after substitution
\[
x(s)=6M,\quad h=1,\quad C=3\sqrt(2)M
\]
 takes a form
 \[
{\frac {d^{2}}{d{s}^{2}}}Q(s)+{\frac {1}{108}}\,{\frac {Q(s)}{{M}^{2}}
}=0
\]
and its solution is
\begin{equation}\label{dryuma:eq22}
Q(s)={\it C_1}\,\cos\left(1/18\,{\frac {\sqrt {3}s}{M}}\right)+{C_2}\,\sin\left(
1/18\,{\frac {\sqrt {3}s}{M}}\right).
\end{equation}

    At last the expression for coordinate $V(s)$ in considered case can be found from the
relation (\ref{dryuma:eq17}).

 It has the form
\begin{equation}\label{dryuma:eq23}
V(s)=-1/9\,A\left (39+i\sqrt {303}\right )\cos\left({\frac {1}{72}}\,{\frac {
\sqrt {3+3\,i\sqrt {303}}s}{M}}\right){\frac {1}{\sqrt {3+3\,i\sqrt {303}}}}+
\end{equation}
\[
+1/3\,s+2/3\,\mu.
\]

    So the formulaes (\ref{dryuma:eq20},\ref{dryuma:eq21},\ref{dryuma:eq22},\ref{dryuma:eq23})
describe the relation between the properties of motion of the test particle on the orbit $x(s)=6M$
in basic physical space with coordinates $(x,\theta,\phi,t)$ and its map
into additional space with coordinates $(P,Q,U,V)$ .

   The solution of the equation (\ref{dryuma:eq21}) relatively parameter $s$ is
\[
s=72\,M \arccos\left({\frac {U \sqrt {3+3\,i\sqrt {303}}\sqrt {2}}{2AM (
(39+i\sqrt {303})}}\right){\frac {1}{\sqrt {3+3\,i\sqrt {303}}}}.
\]

The substitution of this value into the formulae for the coordinate $P(s)$ give us the quadric
\[
95\,i{U}^{2}\sqrt {303}+2071\,{U}^{2}-184832\,{A}^{2}{M}^{2}+184832\,{
P}^{2}{M}^{2}+700416\,P{M}^{3}+663552\,{M}^{4}=0.
\]

In the case of radial motion $C=0,\quad h=1$ we get
\begin{equation}\label{dryuma:eq24}
x(s)=1/2\,{2}^{2/3}{3}^{2/3}\sqrt [3]{M}{s}^{2/3},
\end{equation}
and
\begin{equation}\label{dryuma:eq25}
{\frac {d}{ds}}t(s)=3\,{\frac {{s}^{2/3}}{3\,{s}^{2/3}-2\,{M}^{2/3}
\sqrt [3]{2}\sqrt [3]{3}}}.
\end{equation}

    The system of equations for additional coordinates takes the form
\[
{\frac {d^{2}}{d{s}^{2}}}P(s)+4\,{M}^{2}\sqrt {2}{\frac {d}{ds}}P(s)
\left (x(s)\right )^{-2}{\frac {1}{\sqrt {{\frac {M}{x(s)}}}}}\left (-
2\,M+x(s)\right )^{-1}-\]\[-\left (2\,x(s)\sqrt {{\frac {M}{x(s)}}}+6\,M
\sqrt {{\frac {M}{x(s)}}}\right )MP(s){\frac {1}{\sqrt {{\frac {M}{x(s
)}}}}}\left (-2\,M+x(s)\right )^{-1}\left (x(s)\right )^{-3}-\]\[-{\frac {M
}{x(s)\left (-2\,M+x(s)\right )}}=0,
\]
\[
{\frac {d^{2}}{d{s}^{2}}}Q(s)-2\,{\frac {\left (\left ({\frac {d}{ds}}
x(s)\right )\left (x(s)\right )^{4}-2\,M\left ({\frac {d}{ds}}x(s)
\right )\left (x(s)\right )^{3}\right ){\frac {d}{ds}}Q(s)}{\left (-2
\,M+x(s)\right )\left (x(s)\right )^{4}}}-\]\[-2\,{\frac {\left (-M
\left ({\frac {d}{ds}}t(s)\right )^{2}(x(s)-2\,M)^2-\left ({\frac {d}{ds}}x(s)
\right )^{2}\left (x(s)\right )^{2}(x(s)-3\,M)\right )Q(s)}{\left (-2\,M+x(s)\right )\left (x(s)\right )
^{4}}}=0,
\]
\[
{\frac {d^{2}}{d{s}^{2}}}U(s)+4/3\,{\frac {U(s)}{{s}^{2}}}-4/3\,{
\frac {{\frac {d}{ds}}U(s)}{s}}=0,
\]
and
\[
V(s)+1/2\,{\frac {2\,\left ({\frac {d}{ds}}x(s)\right )P(s)-s-2\,\mu}{
{\frac {d}{ds}}t(s)}}=0.
\]

After substitution here the relations (\ref{dryuma:eq24},\ref{dryuma:eq25}) we find the solutions
\[
P(s)=1/2\,{\frac {\sqrt [3]{s}\sqrt [3]{2}\sqrt [3]{3}\left (3\,C_5+s\sqrt [3
]{2}\sqrt [3]{3}{M}^{2/3}\right )}{-3\,\sqrt [3]{M}{s}^{2/3}+2\,M
\sqrt [3]{2}\sqrt [3]{3}}},
\]
\[
Q(s)={C_3}\,s+{C_4}\,{s}^{4/3},
\]
\[
U(s)={C_1}\,s+{C_2}\,{s}^{4/3},
\]
and
\[
V(s)=s/2+{\frac {C_5}{{s}^{2/3}}}.
\]

  The linear system of geodesics for additional coordinates (\ref{dryuma:eq14})
   may be used for the studying of the  properties of a basic space. In particular the sequence of the matrixes
\[
E(s),\quad E_{;s},\quad E_{;ss},...,
\]
where
\[
E_{s}=\frac{d E(s)}{ds}+\frac{1}{2}\left[A(s),E(s)\right]
\]
and their invariants are important characteristic of a basic space.

Remark that for a given example the matrix $E(s)$ has a property
\[
Det(E(s))=0,\quad Trace(E(s))=0.
\]

More detail consideration leads to conclusion that in general case
for the matrix $E(s)$ the condition
\[
Trace(E(s))=R_{ij} \dot x^{i} \dot x^{j}
\]
is obeyed, where $R_{ij}$ is the Ricci tensor of the basic space.

The generalization and the interpretation of these results will be
done later.

\subsection*{Acknowledgement}

 The author thanks INTAS-99-01782 Programm and the Royal Swedish Academy of Sciences for financial support.

\end{document}